\documentclass[12pt]{article}
\usepackage{epsfig,multicol,latexsym}
\newdimen\digitwidth
\setbox0=\hbox{\rm0}
\digitwidth=\wd0
\def\indnt{\vskip0pt\noindent\vsp\hskip1.5em}
\def\vsp{\hbox{\vrule height12.5pt depth3.5pt width0pt}}
\def\sqr#1#2{{\vcenter{\vbox{\hrule height .#2pt
        \hbox{\vrule width .#2pt height#1pt\kern#1pt
                   \vrule width.#2pt}\hrule height.#2pt}}}}

\def\brlist{\begin{thebibliography}{999}}
\def\brf{\bibitem}
\def\erlist{\end{thebibliography}}

\catcode`@=11
\def\eqalign#1{\null\,\vcenter{\openup\jot\m@th
  \ialign{\strut\hfil$\displaystyle{##}$&$\displaystyle{{}##}$\hfil
      \crcr#1\crcr}}\,}
\def\eqalignno#1{\displ@y \tabskip\@centering
  \halign to\displaywidth{\hfil$\@lign\displaystyle{##}$\tabskip\z@skip
    &$\@lign\displaystyle{{}##}$\hfil\tabskip\@centering
    &\llap{$\@lign##$}\tabskip\z@skip\crcr
    #1\crcr}}
\def\leqalignno#1{\displ@y \tabskip\@centering
  \halign to\displaywidth{\hfil$\@lign\displaystyle{##}$\tabskip\z@skip
    &$\@lign\displaystyle{{}##}$\hfil\tabskip\@centering
    &\kern-\displaywidth\rlap{$\@lign##$}\tabskip\displaywidth\crcr
    #1\crcr}}

\long\def\@makefntext#1{
  \vskip0pt\parindent0pt\begin{list}{}%
  {\labelwidth1.5em\leftmargin=\labelwidth%
     \labelsep3pt\itemsep0pt\parsep0pt\topsep-2pt
            \def\baselinestretch{1.0}\footnotesize}%
  \item[\hfill\@makefnmark]#1\end{list}}

\long\def\@makecaption#1#2{\vskip10pt
    {#1:\ \begingroup\small\baselineskip14pt plus4pt minus2pt
             #2\par\endgroup}}

\newcounter{eqltr}
\def\eqlbl#1{\addtocounter{eqltr}{1}
             \edef\theequation{\theequation\mbox{$\alph{eqltr}$}}
               \let\@currentlabel=\theequation\label{#1}
                (\theequation)}

\def\mylbl#1{\let\@currentlabel=\theequation\label{#1}
                (\theequation)}

\catcode`@=12
\def\bqt#1#2\eqt{\begin{equation}\label{#1}%
        {#2}\end{equation}\vskip3mm\noindent}
\def\bln#1#2\eln{\begin{equation}\label{#1}%
   \eqalign{#2}\end{equation}\vskip3mm\noindent}
\begin{document}
\thispagestyle{empty}
\begin{center}{\sf BNL PREPRINT}\end{center}
\vglue 2mm
\rightline{BNL-QGS-01-0501}
\vskip 16mm
\begin{center}
{\Large\bf $SU(3)$ Classification 
         of $p$-Wave $\eta\pi$ and $\eta'\pi$ Systems\\[3mm]
}
\vskip 8mm
S.~U.~Chung 
\vskip2mm
{\em Physics Department\\[-1mm]
     Brookhaven National Laboratory,$^{*}$ 
Upton, Long Island, NY 11973, U.S.A.}
\vskip 8mm
E.~Klempt
\vskip 2mm
{\em Helmholtz-Institut f\"ur Strahlen- und Kernphysik\\
 Universit\"at Bonn, Nu\ss allee 12, D-53115 Bonn, Germany}
\vskip 8mm
J.G.~K\"orner
\vskip 2mm
{\em Institut f\"ur Physik\\
 Universit\"at Mainz, Staudingerweg 7, D-55099 Mainz, Germany}
\vskip15mm
\today
\vskip18mm

{\large\bf Abstract}
\vskip 1cm
\begin{quote}
\indnt
   An exotic meson, the $\pi_1(1400)$ with $J^{PC}=1^{-+}$,
has been seen to decay into a $p$-wave $\eta\pi$ system.  
If this decay conserves flavor $SU(3)$,
then it can be shown that this exotic meson must be a
four-quark state ($q\bar q+q\bar q$) belonging to 
a flavor ${\bf10}\oplus{\bf\overline{10}}$
representation of $SU(3)$.  In contrast, the  $\pi_1(1600)$
with a substantial decay mode into $\eta'\pi$ is likely to
be a member of a flavor octet.
\end{quote}
\end{center}
\vspace*{\fill}\footnoterule\vskip 2mm\footins{\small$^*$
 under contract number DE-AC02-76CH00016 
                  with the U.S. Department of Energy}
\eject
\pagenumbering{arabic}
\def\baselinestretch{1.4}\normalsize
%
\indnt
An exotic meson, the $\pi_1(1400)$ with $I^G(J^{PC})=1^-(1^{-+})$,
has been seen to decay into a $p$-wave $\eta\pi$ system~\cite{E852a}, 
\cite{CBa}. 
So far this is the only decay mode known for the exotic meson.
The purpose of this letter is to show that a $p$-wave 
$\eta\pi$ system belongs to a flavor ${\bf10}\oplus{\bf\overline{10}}$
representation by the requirement of Bose symmetrization,
if the $\eta$ meson is assumed to be a pure member of the pion octet. 
This implies that, in the limit of flavor $SU(3)$ conservation
in its decay, $\pi_1(1400)$ cannot be a gluonic hybrid
($q\bar q+g$) but instead it must belong to a family of four-quark
states ($q\bar q+q\bar q$). In this letter we give, for the first time,
a complete set of normalized wave functions corresponding to the family of
${\bf10}\oplus{\bf\overline{10}}$ coupling to
two members of the ground-state $^1S_0$.
We then explore the consequences 
of this assignment and suggest experiments which could support
or negate this conjecture.
\par
We start the discussion by recalling that a 
second exotic meson,
the $\pi_1(1600)$ with $I^G(J^{PC})=1^-(1^{-+})$, is reported to have
a substantial decay mode into $\eta'\pi$~\cite{VES1}, \cite{VES2},
\cite{E852b}. 
The branching ratios are~\cite{VES2}
\bln{br0}
   B[b_1(1235)\pi]\ :\ B[\eta'(958)\pi]\ :\ B[\rho(770)\pi]
        =1\ :\ 1.0\pm0.3\ :\ 1.5\pm0.5
\eln
A search for a $\eta\pi$ decay mode of the $\pi_1(1600)$ has  not been
successful.
\par
The first immediate question is: why does the  $\pi_1(1600)$ 
not decay into $\eta\pi$ even though it decays into $\eta'\pi$ 
and even though the phase space for the former decay mode is much larger. 
Sometimes it is argued that the strong coupling to $\eta'\pi$
is due to the ``gluish'' nature of the $\eta'$ and points to
a hybrid interpretation of the  $\pi_1(1600)$.  
We do not share this view. Instead, we show that a meson
belonging to a flavor octet cannot decay into $\eta\pi$. 
\par
First we assume that the $\eta$ meson is the pure 
isoscalar member of the ground-state $^1S_0$ flavor-$SU(3)$ octet, 
i.e. the other members are the isovector $\pi$
and the isodoublets $K$ and $\bar K$.  In this scheme then, 
one has assumed that the $\eta'$ is a pure $SU(3)$ singlet.  
The exotic quantum numbers enforce the $\eta\pi$ system to be
in a relative $p$ wave; hence the $\eta\pi$ system must belong  
to the antisymmetric octet ${\bf8_2}$. This is, however, not possible.
To show this, one needs to examine its nonstrange neutral members\cite{ch1}, 
as given in Table I (for a comment on the particle names used, see
the text following the introduction of Table III).
\vskip3mm
\begin{center}
\vbox{
\catcode`?=\active
\def?{\kern\digitwidth}
\def\arraystretch{1.3}
\begin{tabular}{|l|ccr|r|l|l|}
\hline
\multicolumn{7}{|c|}{Table I: Antisymmetric Octet (${\bf8_2}$)}\\
\hline\hline
$J^{PC}$&$S$&$I$&$I_3$&$Q$&Name&\multicolumn{1}{c|}{wave functions}\\
\hline
$1^{--}$&$0$&1&$+1$&$+1$&?$\sigma_-^+$
        &$\sqrt{1\over3}\,\left(\pi^+\,\pi^0-\pi^0\,\pi^+\right)
        -\sqrt{1\over6}\,\left(\bar K^0\,K^+-K^+\,\bar K^0\right)$\\
  &&&$0$&$0$&?$\sigma_-^0$
        &$\sqrt{1\over3}\,\left(\pi^+\,\pi^--\pi^-\,\pi^+\right)
        -\sqrt{1\over12}\left(\bar K^0\,K^0-K^0\,\bar K^0\right)$\\
          &&&&&&\hspace{10mm}$-\sqrt{1\over12}\left(K^-\,K^+-K^+\,K^-\right)$\\
  &&&$-1$&$-1$&?$\sigma_-^-$
        &$\sqrt{1\over3}\,\left(\pi^0\,\pi^--\pi^-\,\pi^0\right)
        -\sqrt{1\over6}\,\left(K^-\,K^0-K^0\,K^-\right)$\\
\hline
$1^{--}$&$0$&0&0&$0$&?$\lambda_-^0$
        &${1\over2}\,\left(\bar K^0\,K^0-K^0\,\bar K^0\right)
                -{1\over2}\,\left(K^-\,K^+-K^+\,K^-\right)$\\
\hline
\end{tabular}
}
\end{center}
\vskip11mm\noindent 
It is clear that they come with an eigenvalue of $C=-1$, 
and so they cannot couple to $\pi\eta$.
Another way to explain this is to note that, for an antisymmetric $\pi\eta$
to belong to an octet, the wave function for
its isosinglet partner must also be antisymmetric, 
but such a wave function can only come with
$C=-1$ as shown in Table I.  An antisymmetric $\pi\eta$ can appear
in the ${\bf10}\oplus{\bf\overline{10}}$ representation, 
because it is devoid of an isosinglet member.
\indnt
    A summary of the various $J^{PC}$'s allowed for all the
representations in $8\otimes 8$ is given in Table II.
\par
We conclude that the absence of the $\eta\pi$ decay mode of the 
$\pi_1(1600)$ is due to its being a member of an octet of flavor 
$SU(3)$. This conclusion holds, whatever its constituent quark 
content may be, hybrid or four-quark system.
\par
Now we allow for deviations from ideal mixing in the pseudoscalar
meson nonet. Then, the $\pi_1(1400)$ has a small but finite chance
to decay into $\eta'\pi$ and the $\pi_1(1600)$ to decay into
$\eta\pi$. The mixing is, however, not large enough to reverse
the experimental observation of a large coupling of the 
$\pi_1(1400)$ to $\eta\pi$ and of the $\pi_1(1600)$ to 
$\eta'\pi$.
\par
\def\arraystretch{1.5}
\begin{center}
\begin{tabular}{|l|l|l|}
\hline
\multicolumn{3}{|c|}{Table II: $SU(3)$ Multiplets and their Composition }\\
\hline\hline
$SU(3)$ Multiplet & $J^{PC}$ or
$J^{P}$ & Composition\\
\hline\hline
  Singlet ({\bf1}) & ${\rm even}^{++}$ 
        &$q\bar q$, $q\bar q+g$, $q\bar q+q\bar q$\\
\hline
  Symmetric Octet (${\bf8_1}$) & ${\rm even}^{++}$
        &$q\bar q$, $q\bar q+g$, $q\bar q+q\bar q$\\
  Antisymmetric Octet (${\bf8_2}$) & ${\rm odd}^{--}$
        &$q\bar q$, $q\bar q+g$, $q\bar q+q\bar q$\\
\hline
  multiplet {\bf 20} (${\bf10}\oplus{\bf\overline{10}}$) &${\rm odd}^{-}$
        &$q\bar q+q\bar q$ (14 strange states)\\
& ${\rm odd}^{-+}$&$q\bar q+q\bar q$ (3 non-strange states)\\
& ${\rm odd}^{--}$&$q\bar q+q\bar q$ (3 non-strange states)\\
\hline
  Multiplet 27 & ${\rm even}^{++}$
        &$q\bar q+q\bar q$\\
\hline
\end{tabular}
\end{center}
\def\arraystretch{1.0}
\vskip3mm\noindent
A system of $\eta\pi$ in a $p$-wave  has been the subject of long-standing
investigation by theorists for a number of years.  The first reference we are
aware of, on this topic, is that of B.-W. Lee {\it et al.}~\cite{LOS}
who pointed out in 1964 that such a system must necessarily belong to
an `icosuplet' (i.e. ${\bf10}\oplus{\bf\overline{10}}$).
H. J. Lipkin~\cite{HJL} and F. E. Close and H. J. Lipkin~\cite{CLa}
pointed out the exotic nature of a $\phi\pi$ and a $p$-wave $\eta\pi$ state
in the same vein.  For further comments on the $p$-wave $\eta\pi$ system,
the reader is referred to the later articles by different authors~\cite{EX0}.

But why does the $\pi_1(1400)$ decay into $\eta\pi$\,? As shown above,
the requirement of Bose symmetry, coupled with standard $SU(3)$ 
isoscalar factors, 
leads to the conclusion that the $p$-wave $\eta\pi$ final state 
is not contained in a flavor octet. We have to use a higher
SU(3) representation.
\par
The product of two $SU(3)$ octets breaks up into\cite{dSw}
\bln{r8a}
        {\bf 8\otimes 8\ =\ 27\ \oplus\ 10\ \oplus\ \overline{10}\ \oplus\ 
                       8_1\ \oplus\ 8_2\ \oplus\ 1}  
\eln
The 27-plet has even spin only, hence we have to use 
the ${\bf10}\oplus{\bf\overline{10}}$ representation.
The appropriate normalized wave functions can be written down using 
the $SU(3)$ isoscalar factors\cite{dSw}; the results are tabulated
in Table III. 
One sees that all the wave functions are antisymmetric under the interchange
of two states.  In order that the Bose symmetrization holds for the entire
wave function, one must demand that the spatial symmetry - orbital angular
momentum - be odd.  We see that a $p$-wave $\pi\eta$ system must belong
to the ${\bf10}\oplus{\bf\overline{10}}$ representation.
\indnt
   One notes that the particle names in Table III are derived from their
counterparts 
in the Baryon sector but with lower-case letters.  Thus a
$I^G(J^{PC})=1^-(1^{-+})$ 
$\pi_1(1400)$ is given the name $\sigma_+(1400)$ whereby emphasis is placed
on the four-quark nature of the object.  A $\rho$ is a generic name for
a state with $I^G(J^{PC})=1^+(1^{--})$ but the name $\sigma_-$ signifies its
four-quark character.   The counterparts to $K^*$ and $\omega$ are
likewise denoted $\xi$ and $\lambda$, respectively.
\vskip1mm
\begin{center}
\vbox{
\catcode`?=\active
\def?{\kern\digitwidth}
\def\arraystretch{1.3}
\begin{tabular}{|l|ccr|r|l|l|}
\hline
\multicolumn{7}{|c|}{Table IIIa:  
        multiplet {\bf 20} (${\bf10}\oplus{\bf\overline{10}}$)---strange members}\\
\hline\hline
$J^{P}$&$S$&$I$&$I_3$&$Q$&Name&\multicolumn{1}{c|}{wave functions}\\
\hline
$1^-$&$+2$&0&0&$+1$&?$\bar\omega^{+}$
&$-\sqrt{1\over 2}\left(K^+\,K^0-K^0\,K^+\right)$\\
\hline
$1^-$&$+1$&3/2&+3/2&$+2$&?$\delta^{++}$
&${1\over\sqrt{2}}\,\left(\pi^+\,K^+-K^+\,\pi^+\right)$\\
    &&&+1/2&$+1$&?$\delta^{+}$
&$\sqrt{1\over 6}\,\left(\pi^+\,K^0-K^0\,\pi^+\right)
                +\sqrt{1\over 3}\,\left(\pi^0\,K^+-K^+\,\pi^0\right)$\\
    &&&-1/2&$0$&?$\delta^{0}$
&$\sqrt{1\over 3}\,\left(\pi^0\,K^0-K^0\,\pi^0\right)
                +\sqrt{1\over 6}\,\left(\pi^-\,K^+-K^+\,\pi^-\right)$\\
    &&&-3/2&$-1$&?$\delta^{-}$
&${1\over\sqrt{2}}\,\left(\pi^-\,K^0-K^0\,\pi^-\right)$\\
\hline
$1^-$&$+1$&1/2&+1/2&$+1$&?$\bar\xi^{+}$
&$-\sqrt{1\over 6}\,\left(\pi^+\,K^0-K^0\,\pi^+\right)
                +\sqrt{1\over 12}\,\left(\pi^0\,K^+-K^+\,\pi^0\right)$\\
    &&&&&&\hspace{15mm}$+{1\over 2}\,\left(K^+\,\eta-\eta\,K^+\right)$\\
    &&&-1/2&$0$&?$\bar\xi^{0}$
&$-\sqrt{1\over 12}\,\left(\pi^0\,K^0-K^0\,\pi^0\right)
                +\sqrt{1\over 6}\,\left(\pi^-\,K^+-K^+\,\pi^-\right)$\\
    &&&&&&\hspace{15mm}$+{1\over 2}\,\left(K^0\,\eta-\eta\,K^0\right)$\\
\hline
$1^-$&$-1$&1/2&+1/2&?$0$&?$\xi^{0}$
&$\sqrt{1\over 6}\,\left(\pi^+\,K^--K^-\,\pi^+\right)
                -\sqrt{1\over 12}\,\left(\pi^0\,\bar K^0-\bar K^0\pi^0\right)$\\
    &&&&&&\hspace{15mm}$+{1\over 2}\,\left(\bar K^0\,\eta-\eta\,\bar K^0\right)$\\
   &&&-1/2&$-1$&?$\xi^{-}$
&$\sqrt{1\over 12}\,\left(\pi^0\,K^--K^-\,\pi^0\right)
                -\sqrt{1\over 6}\,\left(\pi^-\,\bar K^0-\bar K^0\,\pi^-\right)$\\
    &&&&&&\hspace{15mm}$+{1\over 2}\,\left(K^-\,\eta-\eta\,K^-\right)$\\
\hline
$1^-$&$-1$&3/2&+3/2&$+1$&?$\bar\delta^{+}$
&${1\over\sqrt{2}}\,\left(\pi^+\,\bar K^0-\bar K^0\,\pi^+\right)$\\
     &&&+1/2&$0$&?$\bar\delta^{0}$
&$\sqrt{1\over 6}\,\left(\pi^+\,K^--K^-\,\pi^+\right)
                +\sqrt{1\over 3}\,\left(\pi^0\,\bar K^0-\bar K^0\,\pi^0\right)$\\
     &&&-1/2&$-1$&?$\bar\delta^{-}$
&$\sqrt{1\over 3}\,\left(\pi^0\,K^--K^-\,\pi^0\right)
                +\sqrt{1\over 6}\,\left(\pi^-\,\bar K^0-\bar K^0\,\pi^-\right)$\\
     &&&-3/2&$-2$&?$\bar\delta^{--}$
&${1\over\sqrt{2}}\,\left(\pi^-\,K^--K^-\,\pi^-\right)$\\
\hline
$1^-$&$-2$&0&0&$-1$&?$\omega^{-}$
&$\sqrt{1\over 2}\left(\bar K^0\,K^--K^-\,\bar K^0\right)$\\
\hline
\end{tabular}
}
\end{center}
\begin{center}
\vbox{
\catcode`?=\active
\def?{\kern\digitwidth}
\def\arraystretch{1.3}
\begin{tabular}{|l|ccr|r|l|l|}
\hline
\multicolumn{7}{|c|}{Table IIIb:  
        multiplet {\bf 20} (${\bf10}\oplus{\bf\overline{10}}$)---non-strange members}\\
\hline\hline
$J^{PC}$&$S$&$I$&$I_3$&$Q$&Name&\multicolumn{1}{c|}{wave functions}\\
\hline
$1^{-+}$&$0$&1&+1&$+1$&?$\sigma_+^{+}$
&${1\over\sqrt{2}}\,\left(\pi^+\,\eta-\eta\,\pi^+\right)$\\
  &&&0&$0$&?$\sigma_+^{0}$
&${1\over\sqrt{2}}\,\left(\pi^0\eta-\eta\pi^0\right)$\\
  &&&-1&$-1$&?$\sigma_+^{-}$
&${1\over\sqrt{2}}\,\left(\pi^-\,\eta-\eta\,\pi^-\right)$\\
\hline
$1^{--}$&$0$&1&+1&+1&?$\sigma_-^{+}$
&$\sqrt{1\over 6}\left(\pi^+\,\pi^0-\pi^0\,\pi^+\right)
          +\sqrt{1\over 3}\,\left(\bar K^0\,K^+-K^+\,\bar K^0\right)$\\
  &&&0&0&?$\sigma_-^{0}$
&$\sqrt{1\over 6}\left(\pi^+\,\pi^--\pi^-\,\pi^+\right)
          +\sqrt{1\over 6}\left(\bar K^0\,K^0-K^0\,\bar K^0\right)$\\
       &&&&&&\hspace{10mm}$+\sqrt{1\over 6}\left(K^-\,K^+-K^+\,K^-\right)$\\
  &&&-1&-1&?$\sigma_-^{-}$
&$\sqrt{1\over 6}\left(\pi^0\,\pi^--\pi^-\,\pi^0\right)
          +\sqrt{1\over3}\,\left(K^-\,K^0-K^0\,K^-\right)$\\
\hline
\end{tabular}
}
\vskip1mm
\end{center}
\indnt
   It can be shown that
antisymmetric wave functions (under the interchange of the two
particles) occur for the ${\bf8_2}$, ${\bf10}$ and ${\bf\overline{10}}$ 
representations, all others being symmetric.  
It is obvious that a $p$-wave $\pi\eta$ system  
belongs to the ${\bf10}\oplus{\bf\overline{10}}$ representation, 
as shown in Table III.  
\indnt
Let us now discuss the consequences of a assignement of the
$\pi_1(1400)$ to the ${\bf10}\oplus{\bf\overline{10}}$
multiplet. 
\begin{itemize}
\item The multiplet {\bf 20} does not contain an
isosinglet. This has to be contrasted to the octet
$\pi(1600)$ for which we expect an isosinglet 
partner $\eta_1(1600)$. The $\eta_1(1600)$ could be searched for
in four-pion final states like $a_1(1260)+\pi$ via S-wave or 
$\pi(1300)+\pi$ via P-wave.
\item There are no non-strange mesons with charge 2. 
A possible $Q=\pm2$ candidate has been reported
in the $\pi\pi$ channel\cite{OBa} which, if confirmed, would constitute the first
example of the multiplet ${\bf 27}$.
\item The multiplet {\bf 20} contains mesons with strangeness 2.
They could be searched for in reactions like
\bln{br2}
K^+\, p \rightarrow K^+\, K^0_S\, \Sigma^+  
\eln
The negative strangeness of the $\Sigma^+$ identifies the
$K^0_S$ as $K^0$.
The state with strangeness $S=-2$ is much more difficult
to find since the $K^-K^0_S$ has a 
$S=0$ and a $S=2$ component.
\item The multiplet {\bf 20} contains strange mesons with isospin
3/2. The maximal third component has $I_3=3/2$ and mesons   
with 2 units of charge are supposed to exist.
The search for such particles again requires K$^+$ beams:
\bln{br3}
K^+\, p \rightarrow K^+\, \pi^+\, \Lambda 
\eln
\end{itemize}
We conclude that---in the limit 
in which the $\eta$ is a pure $SU(3)$ octet and the 
$\eta'$ is a pure singlet, and assuming that flavor $SU(3)$
is conserved in the decay---the exotic $\pi_1(1400)$ state 
with quantum numbers $I^G(J^{PC})=1^-(1^{-+})$ and 
seen to decay into $\eta\pi$ and not
into $\eta'\pi$ must be a four-quark
state belonging to the multiplet {\bf 20}, a ${\bf10}\oplus{\bf\overline{10}}$
representation of flavor $SU(3)$, because of the
requirement of Bose symmetrization on a $p$-wave $\pi\eta$ system.
The $\pi_1(1600)$ 
is seen to decay primarily into $\eta'\pi$
and not into $\eta\pi$. This state must belong to an $SU(3)$ octet
and could thus be an exotic meson with a valence gluon in it, 
i.e. $(q\bar q+g)$, although it could just as well be a four-quark state.
A verification of new meson configuratons, in particular 
the verification of mesons in $SU(3)$ decuplets requires 
an intense positively charged Kaon beam. Such experiments 
could be carried out at the RF-separated $K^\pm$ beam currently at
IHEP/Protvino, the 50-GeV Japan Hadron Facility at KEK or at 
the GSI-SIS200 project. 

   Achasov and Shestakov\cite{AS0} has recently worked out the nature of
exotic resonances with $I^G(J^{PC})=1^-(1^{-+})$ in the process
$\rho\pi\to\eta\pi$.  They derive a set of 
unitarized and analytic amplitudes for the process in which the $s$-channel
loop diagrams contain the intermediate states $\rho\pi$, $\eta\pi$,
$\eta'\pi$ and $K^*\bar K+\bar K^*K$.  They consider both 
$\pi_1(1400)$ and $\pi_1(1600)$ in their model and find that an exotic
state coupling to $\eta\pi$ must belong to the 
${\bf10}\oplus{\bf\overline{10}}$ reprsentation of flavor $SU(3)$.
However, their model requires a decay mode $\pi_1(1400)\to\rho\pi$.
We note that---experimentally---the $\pi_1(1400)$ is seen only in the 
$\eta\pi$ decay channel.  If the $\rho$ exchange is absent in the reaction
$\pi^-p\to\pi_1^0(1400)n$, then it can proceed only
via $b_1(1235)$- and $\rho_2(1700)$-Reggeon\cite{rho2} exchanges.
So the intensity of the $\pi_1^0(1400)$  production falls off
as $1/s$ in comparison to that of the $a_2^0(1320)$  
production.
Let $R$ be the ratio of the cross section for
$\pi^-p\to\pi_1^0(1400)n$ to that for $\pi^-p\to a_2^0(1320)n$.  
Then the suggested model predicts
\bln{as0}
R({\rm KEK}):R({\rm BNL}):R({\rm IHEP}):R({\rm CERN})
	=2.7\,:\,1.0\,:0.49\,:\,0.18
\eln
Here the relevant center-of-mass energy squared ($s$) corresponds 
to the $\pi^-p$ interactions at 6.3 GeV/$c$ (KEK\cite{kek}), 
18 GeV/$c$ (BNL\cite{E852a}), 37 GeV/$c$ (IHEP/Protvino\cite{VES1}) 
and 100 GeV/$c$ (CERN\cite{gams}).  Unfortunately, not all of
the requisite cross sections are known experimentally.
\section*{Acknowledgment}
\indnt
   S. U. Chung is indebted to T. L. Trueman for enlightening discussions
and to H. J. Willutzki (now deceased) for his helpful comments
during an early phase in the preparation of this paper.
\brlist
\brf{E852a}  D. R. Thompson {\it et al.}, Phys. Rev. Lett. {\bf 79}, 1630 (1997);
S. U. Chung {\it et al.}, Phys. Rev. D {\bf 60}, 092001 (1999).
\brf{CBa} A. Abele {\it et al.}, Phys. Lett. {\bf B423},  175 (1998);
            A. Abele {\it et al.}, Phys. Lett. {\bf B446},  349 (1999).
\brf{VES1} G. M. Beladidze {\it et al.}, Phys. Lett. {\bf B313}, 276 (1993).
\brf{VES2} A. Zaitsev, Proc. Eighth International Conf. on Hadron
       Spectroscopy, Beijing, China (1999)---edited 
            by W. G. Li, Y. Z. Huang and B. S. Zou, Nucl. Phys. A675, 155c (2000).
\brf{E852b} E. Ivanov {\it et al.}, Phys. Rev. Lett. {\bf 86}, 3977 (2001).
\brf{ch1} Our definition is somewhat at variance from the usual convention
        used for, e.g. $u$ and $d$ quarks.  The reader may wish to consult
        S. U. Chung, 
        ` C- and G-parity: A New Definition and Applications (Version III),'
        on the website http://130.199.22.118/reviews.html. 
        For a non-trivial application, the reader may wish to consult
        S.U. Chung, `Analysis of $K\bar K\pi$ systems,' at the same website.
\brf{LOS} B.-W. Lee, S. Okubo and J. Schechter, Phys. Rev.  {\bf 135}, B219 (1964).
\brf{HJL} H. J. Lipkin, Proc. Second International Conf. on Hadron
       Spectroscopy, KEK, Japan (1987)---edited by Y. Oyanagi {\it et al.}, p. 363.
\brf{CLa} F. E. Close and H. J. Lipkin, Phys. Lett. {\bf 196}, 245 (1987).
\brf{EX0} F. Iddir {\it et al.}, Proc. Symmetry Violations in Subatomic Physics,
                   Kinston, UK (1988), p. 130.
          F. Iddir {\it et al.}, Phys. Lett. B{\bf 205}, 564 (1988);
          F. Iddir {\it et al.}, Phys. Lett. B{\bf 207}, 325 (1988);  
          J. M. Fr\`ere and S. Titard, Phys. Lett. B{\bf 214}, 463 (1988).
\brf{dSw} J. J. de Swart, Rev. Mod. Phys. {\bf 35}, 916 (1963).
\brf{OBa} A. Filippi {\it et al.}, Phys. Lett. B{\bf 495}, 284 (2000).
\brf{AS0} N. N. Achasov and G. N. Shestakov,
	Phys. Rev. D{\bf 63}, 014017 (2000).
\brf{rho2} This trajectory corresponds to the meson $\rho_2(1700)\,[1^3D_2]$, not
	established experimentally but expected in the quark model [see
	S. Godfrey and N. Isgur, Phys. Rev. D{\bf 32}, 189 (1985)].
	It lies `one level' below the leading $\rho$ trajectory which corresponds
	to the mesons $\rho(770)$ and $\rho_3(1690)$.
\brf{kek} H. Aoyagi {\it et al.}, Phys. Lett. B{\bf 314}, 246 (1993).
\brf{gams} D. Alde {\it et al.}, Phys. Lett. B{\bf 205}, 397 (1988).
\erlist
\end{document}